\documentclass[aps,showpacs,preprintnumbers,nofootinbibt,preprint]{revtex4}
\usepackage{graphicx}

\begin{document}

\title{$f\left(R,L_m\right)$ gravity}
\author{Tiberiu Harko}
\email{harko@hkucc.hku.hk}
\affiliation{Department of Physics and
Center for Theoretical and Computational Physics, The University
of Hong Kong, Pok Fu Lam Road, Hong Kong, P. R. China}
\author{Francisco S. N. Lobo}
\email{flobo@cii.fc.ul.pt}
\affiliation{Centro de Astronomia e Astrof\'{\i}sica da
Universidade de Lisboa, Campo Grande, Ed. C8 1749-016 Lisboa,
Portugal}
\begin{abstract}
We generalize the $f(R)$ type gravity models by assuming that the gravitational Lagrangian is given by an arbitrary function of the Ricci scalar $R$ and of the matter Lagrangian $L_m$. We obtain the gravitational field equations in the metric formalism, as well as the equations of motion for test particles, which follow from the covariant divergence of the energy-momentum tensor.  The equations of motion for test particles can also be derived from a variational principle in the particular case in which the Lagrangian density of the matter is an arbitrary function of the energy-density of the matter only. Generally, the motion is non-geodesic, and takes place in the presence of an extra force orthogonal to the four-velocity. The Newtonian limit of the equation of motion is also considered, and a procedure for obtaining the energy-momentum tensor of the matter  is presented. The gravitational field equations and the equations of motion for a particular model in which the action of the gravitational field has an exponential dependence on the standard general relativistic Hilbert--Einstein Lagrange density are also derived.
\end{abstract}


\date{\today}

\maketitle

\section{Introduction}

A very promising way to explain the recent observational data \cite{Ri98,
PeRa03} of the late-time acceleration of the Universe and of dark matter is to assume that at large scales the Einstein gravity model of general relativity
breaks down, and a more general action describes the gravitational field.
Theoretical models in which the standard Einstein-Hilbert action is replaced
by an arbitrary function of the Ricci scalar $R$, first proposed in \cite
{Bu70}, have been extensively investigated lately. Thus, the late-time cosmic acceleration can be explained by $f(R)$ gravity \cite{Carroll:2003wy}, and the conditions of viable cosmological models have been derived in \cite{viablemodels}. In the context of the Solar System regime, severe weak field constraints seem to rule out most of the models proposed so far \cite{solartests,Olmo07}, although viable models do exist \cite{Hu:2007nk,solartests2,Sawicki:2007tf,Amendola:2007nt}. $f(R)$ models that pass local tests and unify inflation with dark energy were considered in \cite{odin}. The possibility that the galactic dynamic of massive test particles can be understood without the need for dark matter was also considered in the framework of $\ f(R)$
gravity models \cite{Cap2,Borowiec:2006qr,Mar1,Boehmer:2007kx,Bohmer:2007fh}. For a review of $f(R)$ generalized gravity models see \cite{SoFa08}.

A generalization of the $f(R)$ gravity theories was proposed in \cite
{Bertolami:2007gv} by including in the theory an explicit coupling of an
arbitrary function of the Ricci scalar $R$ with the matter Lagrangian
density $L_{m}$. As a result of the coupling the motion of the massive
particles is non-geodesic, and an extra force, orthogonal to the
four-velocity, arises. The connections with MOND and the Pioneer anomaly
were also explored. This model was extended to the case of the arbitrary
couplings in both geometry and matter in \cite{ha08}. The astrophysical and cosmological implications of
the non-minimal coupling  matter-geometry coupling were extensively investigated in
\cite{Bertolami:2007vu, ha10}.

It is the purpose of the present paper to consider a maximal extension of the Hilbert-Einstein action, by assuming that the gravitational field can be described by field equations that can be derived from a variational principle with the Lagrangian density of the field given by an arbitrary function of the Ricci scalar and of the matter Lagrangian density.  We derive the field equations for this model, and obtain the equations of motion of the test particles. Generally, the covariant divergence of the energy-momentum tensor is non-zero, and the motion of test particles is non-geodesic, and takes place in the presence of an extra force, orthogonal to the four-velocity.
The non-geodesic motion, due to the non-minimal couplings present in the model, imply the violation of the equivalence principle,
which is highly constrained by solar system experimental tests
\cite{Faraoni,BPT06}. However, it has been recently reported, from
data of the Abell Cluster A586, that interaction of dark matter
and dark energy does imply the violation of the equivalence
principle \cite{BPL07}. Thus, it is possible to test these models with non-minimal couplings in the context of the violation of the equivalence
principle. It is also important to emphasize that the violation of the equivalence principle is also found as a low-energy feature of
some compactified version of higher-dimensional theories.

The Newtonian limit of the equation of motion is also investigated, and a procedure for obtaining the energy-momentum tensor of the matter  is presented. The gravitational field equations and the equations of motion for a particular model in which the action of the gravitational field has an exponential dependence on the standard general relativistic Hilbert--Einstein Lagrange density, $f\left(R,L_m\right)=\Lambda \exp\left(R/2\Lambda +L_m/\Lambda \right)$,  are also derived. Generally, we assume that our model is valid for all types of matter, including scalar fields, vector fields, or even Yang-Mills fields. Non-minimal models of $f(R)$ gravity, in which the geometric part of the action is  multiplied by the electromagnetic field or the Yang-Mills field Lagrangian, were considered in \cite{odin1}.

The present paper is organized as follows. The field equations of the $f\left(R,L_m\right)$ gravity model are derived in Section~\ref{2}. The equation of motion of test particles, and the Newtonian limit of the equation of motion, are considered in Section~\ref{3}. The construction of the energy-momentum tensor for perfect thermodynamics fluids is presented in Section~\ref{4}. A particular model of $f\left(R,L_m\right)$ gravity, in which the Lagrangian density of the gravitational field has an exponential dependence on the standard Hilbert--Einstein Lagrangian density of general relativity is discussed in Section~\ref{5}. We conclude and discuss our results in Section~\ref{6}.

\section{Field equations of $f\left(R,L_m\right)$ gravity}\label{2}

We assume that the action for the modified theories of gravity  takes the following form
\begin{equation}
S=\int f\left(R,L_m\right) \sqrt{-g}\;d^{4}x~,
\end{equation}
where $f\left(R,L_m\right)$ is an arbitrary function of the Ricci scalar $R$, and of the Lagrangian density corresponding to matter, $L_{m}$. We define the
energy-momentum tensor of the matter as \cite{LaLi}
\begin{equation}
T_{\mu \nu }=-\frac{2}{\sqrt{-g}}\frac{\delta \left( \sqrt{-g}L_{m}\right)}{
\delta g^{\mu \nu }}\,.
\end{equation}

By assuming that the Lagrangian density $L_{m}$ of the matter depends only
on the metric tensor components $g_{\mu \nu }$, and not on its derivatives,
we obtain
\begin{equation}\label{en1}
T_{\mu \nu }=g_{\mu \nu }L_{m}-2\frac{\partial L_{m}}{\partial g^{\mu
\nu }}.
\end{equation}

By varying the action $S$ of the gravitational field with respect to the
metric tensor components $g^{\mu \nu }$ we obtain first
\begin{eqnarray}
\delta S&=&\int \left[ f_{R}\left( R,L_{m}\right) \delta R+f_{L_{m}}\left(
R,L_{m}\right) \frac{\delta L_{m}}{\delta g^{\mu \nu }}\delta g^{\mu \nu }\right.-\nonumber\\
&&\left.\frac{1}{2}g_{\mu \nu }f\left( R,L_{m}\right) \delta g^{\mu \nu }\right]
\sqrt{-g}d^{4}x,
\end{eqnarray}
where we have denoted  $f_{R}\left( R,L_{m}\right) =\partial f\left(
R,L_{m}\right) /\partial R$ and $f_{L_{m}}\left( R,L_{m}\right) =\partial f\left(
R,L_{m}\right) /\partial L_{m}$, respectively. For the variation of the
Ricci scalar we obtain
\begin{equation}
\delta R=\delta \left( g^{\mu \nu }R_{\mu \nu }\right) =R_{\mu \nu }\delta
g^{\mu \nu }+g^{\mu \nu }\left( \nabla _{\lambda }\delta \Gamma _{\mu \nu
}^{\lambda }-\nabla _{\nu }\delta \Gamma _{\mu \lambda }^{\lambda }\right) ,
\end{equation}
where $\nabla _{\lambda }$ is the covariant derivative with respect to the
symmetric connection $\Gamma $ associated to the metric $g$. Taking into
account that the variation of the Christoffel symbols can be written as
\begin{equation}
\delta \Gamma _{\mu \nu }^{\lambda }=\frac{1}{2}g^{\lambda \alpha }\left(
\nabla _{\mu }\delta g_{\nu \alpha }+\nabla _{\nu }\delta g_{\alpha \mu
}-\nabla _{\alpha }\delta g_{\mu \nu }\right) ,
\end{equation}
we obtain for the variation of the Ricci scalar the expression
\begin{equation}
\delta R=R_{\mu \nu }\delta g^{\mu \nu }+g_{\mu \nu }\nabla _{\mu}\nabla ^{\mu }  \delta g^{\mu
\nu }-\nabla _{\mu }\nabla _{\nu }\delta g^{\mu \nu }.
\end{equation}

Therefore for the variation of the action of the gravitational field we
obtain
\begin{eqnarray}
\delta S&=&\int \left[ f_{R}\left( R,L_{m}\right) R_{\mu \nu }\delta g^{\mu
\nu }\right.+\nonumber\\
&&\left.f_{R}\left( R,L_{m}\right) g_{\mu \nu }\nabla _{\mu }\nabla ^{\mu } \delta g^{\mu \nu
}\right.-\nonumber\\
&&\left.f_{R}\left( R,L_{m}\right) \nabla _{\mu }\nabla _{\nu }\delta g^{\mu \nu
}+f_{L_{m}}\left( R,L_{m}\right) \frac{\delta L_{m}}{\delta g^{\mu \nu }}%
\delta g^{\mu \nu }\right.-\nonumber\\
&&\left.\frac{1}{2}g_{\mu \nu }f\left( R,L_{m}\right) \delta
g^{\mu \nu }\right] \sqrt{-g}d^{4}x\,.  \label{var1}
\end{eqnarray}

After partially integrating the second and third terms in Eq.~(\ref{var1}),
and with the use of the definition of the energy-momentum tensor, given by Eq.~(\ref{en1}), we obtain
the field equations of the $f\left( R,L_{m}\right) $ gravity model as
\begin{eqnarray}\label{field}
&&f_{R}\left( R,L_{m}\right) R_{\mu \nu }+\left( g_{\mu \nu }\nabla _{\mu }\nabla^{\mu } -\nabla
_{\mu }\nabla _{\nu }\right) f_{R}\left( R,L_{m}\right) -\nonumber\\
&&\frac{1}{2}\left[
f\left( R,L_{m}\right) -f_{L_{m}}\left( R,L_{m}\right)L_{m}\right] g_{\mu \nu }=\nonumber\\
&&\frac{1}{2}%
f_{L_{m}}\left( R,L_{m}\right) T_{\mu \nu }.
\end{eqnarray}

If $f\left( R,L_{m}\right) =R/2+L_{m}$ (the Hilbert-Einstein Lagrangian), from Eq.~(\ref{field}) we
recover the standard equations of general relativity,  $R_{\mu \nu
}-(1/2)g_{\mu \nu }R=T_{\mu \nu }$. For $f\left( R,L_{m}\right)
=f_{1}(R)+f_{2}(R)G\left( L_{m}\right) $, where $f_{1}$, $f_{2}$ and $G$ are
arbitrary functions of the Ricci scalar and of the matter Lagrangian
density, respectively, we reobtain the field equations of the modified
gravity with arbitrary geometry-matter coupling, considered in \cite{ha08}.

The contraction of Eq.~(\ref{field}) provides the following relation between the Ricci scalar $R$, the matter Lagrange density $L_{m}$, and the trace $T=T_{\mu }^{\mu }$ of the energy-momentum tensor,
\begin{eqnarray}
&&f_{R}\left( R,L_{m}\right) R+3\nabla _{\mu }\nabla ^{\mu } f_{R}\left( R,L_{m}\right) -2\left[
f\left( R,L_{m}\right)\right. -\nonumber\\
&&\left.f_{L_{m}}\left( R,L_{m}\right)L_{m}\right]
=\frac{1}{2}f_{L_{m}}\left(
R,L_{m}\right) T.  \label{contr}
\end{eqnarray}

By eliminating the term $\nabla _{\mu }\nabla ^{\mu } f_{R}\left( R,L_{m}\right) $ between Eqs.~(%
\ref{field}) and (\ref{contr}), we obtain another form of the gravitational
field equations as
\begin{eqnarray}
&&f_{R}\left( R,L_{m}\right) \left( R_{\mu \nu }-\frac{1}{3}Rg_{\mu \nu
}\right) +\nonumber\\
&&\frac{1}{6}\left[ f\left( R,L_{m}\right) -f_{L_{m}}\left( R,L_{m}\right)L_{m}\right] g_{\mu
\nu }=\nonumber\\
&&\frac{1}{2}f_{L_{m}}\left( R,L_{m}\right) \left( T_{\mu \nu }-\frac{1}{%
3}Tg_{\mu \nu }\right) +\nonumber\\
&&\nabla _{\mu }\nabla _{\nu }f_{R}\left(
R,L_{m}\right) .
\end{eqnarray}

By taking the covariant divergence of
Eq.~(\ref{field}), with the use of the mathematical identity \cite{Ko06},
\begin{eqnarray}
&&\nabla ^{\mu }\left[ f_R\left(R,L_m\right)R_{\mu \nu }-\frac{1}{2}f\left(R,L_m\right)g_{\mu \nu }\right.+\nonumber\\
&&\left.\left(g_{\mu \nu }\nabla _{\mu }\nabla ^{\mu } -\nabla _{\mu }\nabla _{\nu }\right) f_R\left(R,L_m\right)\right] \equiv 0\,,
\end{eqnarray}
we obtain for the divergence of the energy-momentum tensor $T_{\mu \nu}$ the equation
\begin{eqnarray}
&&\nabla ^{\mu }T_{\mu \nu }=\nabla ^{\mu }\ln \left[
f_{L_m}\left(R,L_m\right)\right] \left\{ L_{m}g_{\mu \nu
}-T_{\mu \nu }\right\} =\nonumber\\
&&2\nabla
^{\mu }\ln \left[ f_{L_m}\left(R,L_m\right) \right] \frac{\partial L_{m}}{%
\partial g^{\mu \nu }}\,.  \label{noncons}
\end{eqnarray}

The requirement of the conservation of the energy-momentum tensor
of matter, $\nabla ^{\mu }T_{\mu \nu }=0$, gives an effective
functional relation between the matter Lagrangian density and the function $f_{L_m}\left(R,L_m\right)$,
\begin{equation}
\nabla
^{\mu }\ln \left[ f_{L_m}\left(R,L_m\right) \right] \frac{\partial L_{m}}{%
\partial g^{\mu \nu }}=0\,.
\end{equation}

Thus, once the matter Lagrangian density is known, by an
appropriate choice of the function $f\left( R,L_{m}\right) $ one can construct, at least in principle, conservative
models with arbitrary matter-geometry dependence.

\section{The equation of motion of test particles and the Newtonian limit in $f\left(R,L_m\right)$ gravity}\label{3}

As a specific example of $f\left(R,L_m\right)$ gravity model we consider the case in which the matter, assumed to be a perfect thermodynamic fluid, obeys a barotropic equation of state, with the thermodynamic pressure $p$ being a function of the {\it rest mass density of the matter} (for short: {\it matter density})  $\rho $ only, so that $p=p\left( \rho \right) $. In this case, the matter Lagrangian density, which in the general case could be a function of both density and pressure, $L_{m}=L_{m}\left( \rho ,p\right) $, or of only one of the thermodynamic parameters, becomes an arbitrary function of the density of the matter $\rho $ only, so that $L_{m}=L_{m}\left( \rho \right) $. Then the energy-momentum tensor of the matter is given by \cite{ha08}
\begin{equation}
T^{\mu \nu }=\rho \frac{dL_{m}}{d\rho }u^{\mu }u^{\nu }+\left( L_{m}-\rho
\frac{dL_{m}}{d\rho }\right) g^{\mu \nu }\,,  \label{tens}
\end{equation}
where the four-velocity $u^{\mu }=dx^{\mu }/ds$ satisfies the condition $%
g^{\mu \nu }u_{\mu }u_{\nu }=1$. To obtain Eq.~(\ref{tens}) we have imposed
the condition of the conservation of the matter current, $\nabla _{\nu
}\left( \rho u^{\nu }\right) =0$, and we have used the relation
\begin{equation}
\delta \rho
=\frac{1}{2} \rho \left( g_{\mu \nu }-u_{\mu }u_{\nu }\right) \delta
g^{\mu \nu },
\end{equation}
whose proof is given in the Appendix of \cite{ha10}. With the use of the
identity
\begin{equation}
u^{\nu }\nabla _{\nu }u^{\mu }=\frac{d^{2}x^{\mu }}{ds^{2}}+\Gamma _{\nu
\lambda }^{\mu }u^{\nu }u^{\lambda },
\end{equation}
from Eqs.~(\ref{noncons}) and (\ref
{tens}) we obtain the equation of motion of a test fluid in $f\left(R,L_m\right)$ gravity as
\begin{equation}
\frac{d^{2}x^{\mu }}{ds^{2}}+\Gamma _{\nu \lambda }^{\mu }u^{\nu }u^{\lambda
}=f^{\mu },  \label{eqmot}
\end{equation}
where
\begin{equation}
f^{\mu }=-\nabla _{\nu }\ln \left[  f_{L_m}\left(R,L_m\right) \frac{%
dL_{m}\left( \rho \right) }{d\rho }\right] \left( u^{\mu }u^{\nu }-g^{\mu
\nu }\right) .
\end{equation}
The extra-force $f^{\mu }$  is perpendicular to the four-velocity, $f^{\mu
}u_{\mu }=0$.

The equation of motion Eq.~(\ref{eqmot}) can be obtained from the variational principle
\begin{equation}
\delta S_{p}=\delta \int L_{p}ds=\delta \int \sqrt{Q}\sqrt{g_{\mu \nu
}u^{\mu }u^{\nu }}ds=0\,,  \label{actpart}
\end{equation}
where $S_{p}$ and $L_{p}=\sqrt{Q}\sqrt{g_{\mu \nu }u^{\mu }u^{\nu }}$ are
the action and the Lagrangian density for test particles, respectively,
and
\begin{equation}
\sqrt{Q}=f_{L_m}\left(R,L_m\right) \frac{dL_{m}\left( \rho \right) }{%
d\rho }\,.  \label{Q}
\end{equation}

To prove this result we start with the Lagrange equations corresponding to
the action~(\ref{actpart}),
\begin{equation}
\frac{d}{ds}\left( \frac{\partial L_{p}}{\partial u^{\lambda }}\right) -%
\frac{\partial L_{p}}{\partial x^{\lambda }}=0.
\end{equation}

Since
\begin{equation}
\frac{\partial L_{p}}{\partial u^{\lambda }}=\sqrt{Q}u_{\lambda }
\end{equation}
and
\begin{equation}
 \frac{\partial L_{p}}{\partial x^{\lambda }}=\frac{1}{2} \sqrt{Q}g_{\mu \nu,\lambda }u^{\mu }u^{\nu }+\frac{ 1}{2} \frac{Q_{,\lambda }}{Q},
\end{equation}
a straightforward calculation gives the equations of motion of the particle as
\begin{equation}
\frac{d^{2}x^{\mu }}{ds^{2}}+\Gamma _{\nu \lambda }^{\mu }u^{\nu }u^{\lambda
}+\left( u^{\mu }u^{\nu }-g^{\mu \nu }\right) \nabla _{\nu }\ln \sqrt{Q}=0.
\end{equation}
By simple identification with the equation of motion  given by Eq.~(\ref{eqmot}), we
obtain the explicit form of $\sqrt{Q}$, as given by Eq.~(\ref{Q}). When $\sqrt{Q}\rightarrow 1$ we reobtain the standard general relativistic equation for geodesic motion.

The variational principle~(\ref{actpart}) can be used to study the Newtonian
limit of the equations of motion of the test particles. In the limit of the weak gravitational fields,
\begin{equation}
ds\approx \sqrt{1+2\phi -\vec{v}^{2}}dt\approx \left( 1+\phi -\vec{v}
^{2}/2\right) dt\,,
\end{equation}
where $\phi $ is the Newtonian potential and $\vec{v}$ is
the usual tridimensional velocity of the fluid.
By assuming that in the Newtonian limit of weak gravitational fields the function $f_{L_m}\left(R,L_m\right)$ can be represented as
\begin{equation}
f_{L_m}\left(R,L_m\right)\approx 1+U\left(R, L_m\right)\,,
\end{equation}
where $U\left(R,L_m\right)\ll 1$, the function $\sqrt{Q}$ can be obtained as
\begin{equation}
\sqrt{Q}=\frac{dL_{m}\left( \rho \right) }{d\rho }+U\left(R,L_m\right)\frac{
dL_{m}\left( \rho \right) }{d\rho }\,.
\end{equation}

Taking into account the first order of approximation, the equations of motion of the fluid can be derived from the variational principle
\begin{equation}
\delta \int \left[ \frac{dL_{m}\left( \rho \right) }{d\rho }+U\left(R,L_m\right)%
\frac{dL_{m}\left( \rho \right) }{d\rho }+\phi -\frac{\vec{v}^{2}}{2}\right]
dt=0\,,
\end{equation}
and are given by
\begin{equation}
\vec{a}=-\nabla \phi -\nabla \frac{dL_{m}\left( \rho \right) }{d\rho }%
-\nabla U_{E}=\vec{a}_{N}+\vec{a}_{H}+\vec{a}_{E}\,,
\end{equation}
where $\vec{a}$ is the total acceleration of the system, $\vec{a}%
_{N}=-\nabla \phi $ is the Newtonian gravitational acceleration, and
\begin{equation}
\vec{a}_{E}=-\nabla U_{E}=-\nabla \left[ U\left(R,L_m\right)\frac{dL_{m}\left( \rho \right)}
{d\rho }\right]\,,
\end{equation}
is a supplementary acceleration induced due to the modification of the action of the gravitational field. As for the term $\vec{a}_{H}=-\nabla \left[
dL_{m}\left( \rho \right) /d\rho \right] $, it has to be identified with the
hydrodynamic acceleration term in the perfect fluid Euler equation.

\section{The matter energy-momentum tensor in $f\left(R,L_m\right)$ gravity}\label{4}

Since $\vec{a}_H$ can be interpreted as a hydrodynamic acceleration term, in the Newtonian limit it can be represented in the form
\begin{equation}
\vec{a}_{H}=-\nabla \frac{dL_{m}\left( \rho \right) }{d\rho }=-\nabla
\int_{\rho _{0}}^{\rho }\frac{dp}{d\rho }\frac{d\rho }{\rho }\,,
\end{equation}
where $\rho _{0}$, an integration constant, plays the role of a limiting
density. Hence the matter Lagrangian can be obtained by a simple integration
as
\begin{equation}
L_{m}\left( \rho \right) =\rho \left[ 1+\Pi \left( \rho \right) \right]
-\int_{p_{0}}^{p}dp\,,  \label{Lm}
\end{equation}
where $\Pi \left( \rho \right) =\int_{p_{0}}^{p}dp/\rho $, and we have normalized an arbitrary integration constant to one. The quantity $p_{0}$ is an integration constant, or a limiting pressure. Therefore the corresponding
energy-momentum tensor of the matter satisfying a barotropic equation of state is given by
\begin{equation}
T^{\mu \nu }=\left\{ \rho \left[ 1+\Phi \left( \rho \right) \right] +p\left(
\rho \right) \right\} u^{\mu }u^{\nu }-p\left( \rho \right) g^{\mu \nu }\,,
\label{tens1}
\end{equation}
respectively, where
\begin{equation}
\Phi \left( \rho \right) =\int_{\rho _{0}}^{\rho }\frac{p}{\rho ^{2}}d\rho
=\Pi \left( \rho \right) -\frac{p\left( \rho \right) }{\rho }\,,
\end{equation}
and with all the constant terms included in the definition of $p$. By introducing the energy density of the massive body according to the definition
\begin{equation}
\varepsilon=\rho \left[ 1+\Phi \left( \rho \right) \right]\,,
 \end{equation}
the energy-momentum tensor of a test fluid can be written in $f\left(R,L_m\right)$ gravity in a form similar to the standard general relativistic case,
\begin{equation}
T^{\mu \nu }=\left[\varepsilon \left(\rho \right)+p\left(\rho \right)\right] u^{\mu }u^{\nu }-p\left( \rho \right) g^{\mu \nu }\,.
\end{equation}

From a physical point of view $\Phi \left( \rho \right) $ can be interpreted
as the elastic (deformation) potential energy of the body, and therefore
Eq.~(\ref{tens1}) corresponds to the energy-momentum tensor of a
compressible elastic isotropic system. An energy-momentum tensor of a similar form was introduced in the framework of standard general relativity by Fock \cite{fock}. The matter Lagrangian can also be
written in the simpler form $L_{m}\left( \rho \right) =\rho \Phi \left( \rho
\right) $.

The form of the matter Lagrangian, and the energy-momentum tensor, are strongly dependent on the equation of state. For example, if the barotropic equation of state is linear, $p=\left( \gamma -1\right) \rho $, $\gamma =$ constant, $1\leq \gamma \leq 2$, then \cite{ha10}
\begin{equation}
L_{m}\left( \rho \right) =\rho \left\{ 1+\left(
\gamma -1\right) \left[  \ln \left( \frac{\rho }{\rho _{0}}\right) -1 %
\right] \right\},
\end{equation}
 and
 \begin{equation}
 \Phi \left( \rho \right) =\left( \gamma -1\right)
\ln \left( \frac{\rho }{\rho _{0}}\right) ,
\end{equation}
 respectively. In the case of a
polytropic equation of state $p=K\rho ^{1+1/n}$, $K,n=$constant, we find
\begin{equation}
L_{m}\left( \rho \right) =\rho +K\left(\frac{n^{2}}{n+1}-1\right)
\rho ^{1+1/n},
\end{equation}
and
\begin{equation}
\Phi \left( \rho \right) =Kn\rho ^{1+1/n}=np\left( \rho
\right) ,
\end{equation}
respectively, where we have taken for simplicity $\rho_{0}=p_{0}=0$. For a fluid satisfying the ideal gas equation of state $p=k_{B}\rho T/\mu $, where $k_{B}$ is Boltzmann's constant, $T$ is the temperature, and $\mu $ is the mean molecular weight, we obtain
\begin{equation}
L_{m}\left( \rho \right) =\rho \left\{ 1+\frac{k_{B}T}{\mu }\left[
\ln \left( \frac{\rho }{\rho _{0}}\right) -1\right] \right\} +p_{0},
\end{equation}
and
\begin{equation}
\Phi \left(\rho \right)=\frac{k_BT}{\mu }\ln\frac{\rho }{\rho _0}.
\end{equation}
respectively. In the case of a physical system satisfying the ideal gas equation of state, the extra-acceleration is given by
\begin{equation}
\vec{a}_{E}\approx - \frac{k_{B}T}{\mu }\nabla \left[ f_{L_m}\left(R,L_m\right) \ln
\frac{\rho }{\rho _{0}}\right] ,
\end{equation}
and it is proportional to the temperature of the fluid.

In $f\left(R,L_m\right)$ gravity the extra-force does not vanish for any specific choices of the matter Lagrangian. In the case of the dust, with $p=0$, the extra force is given by
\begin{equation}
f^{\mu }=-\nabla _{\nu }\ln \left[f_{L_m}\left(R,L_m\right)\right] \left( u^{\mu
}u^{\nu }-g^{\mu \nu }\right) ,
\end{equation}
and it is independent on the thermodynamic properties of the system, being
completely determined by geometry, kinematics and coupling. In the limit of
small velocities and weak gravitational fields, the extra-acceleration of a
dust fluid is given by
\begin{equation}
\vec{a}_{E}=- \nabla \left[ f_{L_m}\left(R,L_m\right)\right].
\end{equation}

If the pressure does not have a thermodynamic or radiative component one can
take $p_{0}=0$. If the pressure is a constant background quantity,
independent of the density, so that $p=p_{0}$, then $L_{m}\left( \rho
\right) =\rho $, and the energy-momentum tensor of the matter takes the form
corresponding to dust, $T^{\mu \nu }=\rho u^{\mu }u^{\nu }$.

\section{Exponential $f\left(R,L_m\right)$ gravity}\label{5}

As a simple toy model for $f\left(R,L_m\right)$ gravity we consider that the gravitational field can be described by a Lagrangian density of the form
\begin{equation}
f\left(R,L_m\right)=\Lambda \exp \left(\frac{1}{2\Lambda }R+\frac{1}{\Lambda }L_m\right),
\end{equation}
where $\Lambda >0$ is an arbitrary constant. In the limit $\left( 1/2\Lambda \right) R+\left( 1/\Lambda \right) L_{m}\ll 1$, we obtain
\begin{equation}
f\left( R,L_{m}\right)  \approx \Lambda +\frac{R}{2}+L_{m}+...,
\end{equation}
that is, we recover the full Hilbert-Einstein gravitational Lagrangian with a cosmological constant.

With this choice of the Lagrangian density the gravitational field equations take the form
\begin{eqnarray}
&&R_{\mu \nu }=\left( \Lambda -L_{m}\right) g_{\mu \nu }+T_{\mu \nu }-
  \nonumber\\
&&\hspace{-0.25cm}\frac{1}{\Lambda }\left[ \left( \frac{1}{2}\nabla _{\mu }\nabla ^{\mu} R+\nabla _{\mu} \nabla ^{\mu } L_{m}\right) g_{\mu \nu}\right.-\nonumber\\
&&\left.\left( \frac{1}{2}\nabla _{\mu }\nabla _{\nu }R+\nabla _{\mu }\nabla _{\nu}L_{m}\right) \right]-
  \nonumber\\
&&\frac{1}{\Lambda ^{2}}\left[ \left( \frac{1}{2}\nabla
^{\lambda }R+\nabla ^{\lambda }L_{m}\right) \left( \frac{1}{2}\nabla
_{\lambda }R+\nabla _{\lambda }L_{m}\right) g_{\mu \nu }\right.-
   \nonumber\\
&&\left.\left( \frac{1}{2}\nabla _{\mu }R+\nabla _{\mu }L_{m}\right) \left( \frac{1}{2}\nabla _{\nu}R+\nabla _{\nu }L_{m}\right) \right] \,.
\end{eqnarray}

By introducing the standard Einstein--Hilbert Lagrangian density of the
gravitational field $L_{H}=R/2+L_{m}$, the gravitational field equations for
the exponential $f\left( R,L_{m}\right) $ gravity model take the form
\begin{eqnarray}\label{exp}
&&R_{\mu \nu }=\left( \Lambda -L_{m}\right) g_{\mu \nu }+T_{\mu \nu }-\nonumber\\
&&\frac{1}{
\Lambda }\left[ \nabla _{\mu }\nabla ^{\mu} L_{H}g_{\mu \nu }-\nabla _{\mu }\nabla _{\nu }L_{H}
\right]- \nonumber\\
&&\frac{1}{\Lambda ^{2}}\left[ \nabla ^{\lambda }L_{H}\nabla
_{\lambda }L_{H}g_{\mu \nu }-\nabla _{\mu }L_{H}\nabla _{\nu }L_{H}\right] .
\end{eqnarray}

In the case of dust, with $L_{m}=\rho $, the mass density of the particles, the equation of motion of test particles becomes
\begin{equation}
\frac{d^{2}x^{\mu }}{ds^{2}}+\Gamma _{\alpha \beta }^{\mu }u^{\alpha
}u^{\beta }=-\frac{1}{\Lambda }\nabla _{\nu }\left( \frac{1}{2}%
R+L_{m}\right) \left( u^{\mu }u^{\nu }-g^{\mu \nu }\right) .
\end{equation}

In the case of weak gravitational fields and of small particle velocities, the extra-acceleration induced in the exponential model is given by $\vec{a}=-\left( 1/\Lambda \right) \left[ \nabla \left( R/2\right) +\nabla
L_{m}\right] $. The extra-acceleration induced by the matter-geometry coupling is proportional to the gradients of the Ricci scalar, and of the matter Lagrangian. In the case of dust, the extra-force is proportional to the gradient of the matter density $\rho $.

As one can see from Eq.~(\ref{exp}), the gravitational field equations of the exponential $f_{L_m}\left(R,L_m\right)$ gravity model contain a background term, proportional to the metric tensor $g_{\mu \nu }$, and which depends on both the constant $\Lambda $, and the physical parameters (density, temperature, pressure) of the matter. Thus the model introduces an effective, time dependent ``cosmological constant''. The presence of such a term could explain the late acceleration of the Universe, which is suggested by the recent astronomical observations \cite{Ri98}.

\section{Discussions and final remarks}\label{6}

In the present paper we have considered a generalized gravity model with a Lagrangian density of an arbitrary function of the Lagrange density of the matter, and of the Ricci scalar. The proposed action represents the most general extension of the standard Hilbert action for the gravitational field,
$S=\int \left[ R/2+L_{m}\right] \sqrt{-g}d^{4}x$. Note that the classical Einstein--Hilbert action, as well as most of its generalizations, has an {\it additive} structure, being constructed as the {\it sum} of several terms describing independently geometry, matter, and their coupling. On the other hand, the $f\left(R,L_m\right)$ modified theories of gravity open the possibility of going beyond this algebraic structure - for example in the exponential $f\left(R,L_m\right)$ gravity model the gravitational Lagrangian density is the {\it product} of two independent functions of the Ricci scalar and of the matter Lagrange density. The equations of motion corresponding to this model show the presence of an extra-force acting on test particles, and the motion is generally non-geodesic. The physical implications of such a force have been already analyzed in the framework of the generalized gravity model with linear and general coupling between matter and geometry, proposed in \cite{Bertolami:2007gv,ha08,Bertolami:2007vu}, and the possible implications for the dark matter problem and for the explanation of the Pioneer anomaly have also been considered.

On the other hand, the field equations Eqs.~(\ref{field}) are equivalent to the Einstein equations of the $f(R)$ model in empty space-time, but differ from them, as well as from standard general relativity, in the presence of matter. Therefore the predictions of the present model could lead to some
major differences, as compared to the predictions of standard general relativity, or its extensions ignoring the role of the matter, in several problems of current interest, like cosmology, gravitational collapse or the generation of gravitational waves. The study of these phenomena may also provide some specific signatures and effects, which could distinguish and discriminate between the various theories of modified gravity.

\section*{Acknowledgments}

We would like to thank to the two anonymous referees, whose comments and suggestions helped us to significantly improve the manuscript. TH is supported by an RGC grant of the government of the Hong Kong SAR. FSNL acknowledges financial support of the Funda\c{c}\~{a}o para a Ci\^{e}ncia e Tecnologia through the grants PTDC/FIS/102742/2008 and CERN/FP/109381/2009.

\end{document}